\documentclass[3p,times,twocolumn]{elsarticle}
\usepackage{ecrc}
\usepackage{amsmath}
\usepackage{braket}
\usepackage{float}
\usepackage[caption=false]{subfig}
\usepackage{dcolumn}
\usepackage{epstopdf}
\usepackage{units}
\usepackage{verbatim}
\usepackage{xcolor}
\usepackage{tabu}
\usepackage{booktabs}
\RequirePackage{xspace}
\usepackage{relsize}
\usepackage{balance}

\volume{00}

\firstpage{1}

\journalname{Nuclear Physics B Proceedings Supplement}

\runauth{}

\jid{nuphbp}

\jnltitlelogo{Nuclear Physics B Proceedings Supplement}

\usepackage{amssymb}
\usepackage[figuresright]{rotating}

\begin{document}
\balance
\begin{frontmatter}



\dochead{}

\title{$D^0$-$\bar{D}^0$ mixing and CP violation results from Belle}


\author{N.K. Nisar}

\address{Tata Institute of Fundamental Research\\Homi Bhabha Road, Colaba, Mumbai 400005, INDIA}

\begin{abstract}
We report the results on $D^0$-$\bar{D}^0$ mixing in the decays of $D^0\to K^+\pi^-$ and $D^0\to K_S^0\pi^+\pi^-$, and CP violation in $D^0\to\pi^0\pi^0$ using a data sample with an integrated luminosity of about 1~ab$^{-1}$ recorded with the Belle detector, at different center of mass energies.  The mixing is observed in $D^0\to K^+\pi^-$ with the mixing parameters $x'^2=(0.09\pm 0.22)\times 10^{-3}$,  $y'=(4.6\pm 3.4)\times 10^{-3}$ and the ratio of doubly Cabibbo-suppressed to Cabibbo-favored decay rates $R_{D}=(3.53\pm0.13)\times 10^{-3}$, where the quoted uncertainties combine both statistical and systematic contributions. For the $D^0\to K_S^0\pi^+\pi^-$ decay, assuming CP conservation we measure mixing  parameters $x=(0.56\pm 0.19^{+0.03+0.06}_{-0.09-0.09})\%$ and $y=(0.30\pm 0.15^{+0.04+0.03}_{-0.05-0.06})\%$, where the uncertainties are statistical, experimental systematics, and that due to amplitude model. We measure the mixing parameters
allowing CP violation for this decay mode, and obtain $|q/p|=0.90^{+0.16+0.05+0.06}_{-0.15-0.04-0.05}$ and arg($q/p$)=$(-6\pm 11 \pm3^{+3}_{-4})^{\circ}$ that are consistent with no CP violation. The time-integrated CP asymmetry in the decay $D^0\to\pi^0\pi^0$ is measured to be, $[-0.03\pm 0.64(stat) \pm 0.10(syst)]\%$, consistent with CP conservation. We also present an updated measurement of CP asymmetry in $D^0\to K_S^0\pi^0$.
\end{abstract}


\end{frontmatter}

%\linenumbers

\section{Introduction}
In the mixing, a pure flavor state of $D^0$ or $\bar{D}^0$ at its production vertex (at time t=0) becomes a mixture of both  $D^0$ and $\bar{D}^0$ in moments later. Time evolution of the neutral $D$ system is described by a non-hermitian Hamiltonian, and its eigenstates $\Ket{D_{H,L}}=p\Ket{D^0}\pm q\Ket{\bar{D}^0}$, are known as mass eigenstates. As evident, the mass eigenstates are a mixture of the flavor eigen states, which causes mixing. This mixing can proceeds either via a short-distance effect involving virtual quarks and $W$ boson, or through a long-distance contribution that involves real kaons or pions in the intermediate state. The phenomenology of mixing is described by two parameters, $x=(m_{H}-m_{L})/\Gamma$ and $y=(\Gamma_{H}-\Gamma_{L})/2\Gamma$, where $m_{H}$, $m_{L}$ are masses and $\Gamma_{H}$, $\Gamma_{L}$ are widths of the mass eigenstates.
These mixing parameters are well measured for $B^{0}$, $B_s^0$ and $K^0$ mesons, but not for $D^0$ mesons.

Complex parameters $p$ and $q$ appearing in the expression of mass eigenstates describe indirect CP violation, with $|p/q|\ne1$ gives CP violation in mixing while $\text{arg}(p/q)\ne 0$ leads to CP violation due to interference between decays with and without mixing. On the other hand, the direct CP violation arises from the difference in amplitudes for particles and their antiparticles decaying to some final state. The phenomenon of CP violation is well established in  $B^{0}$ and $K^0$ systems, while it is not observed in the charm sector.
The contribution of indirect CP violation is mode independent for any neutral mesons, while the direct component is mode dependent. Only time integrated CP asymmetry, a sum of the indirect and direct asymmetries, can be measured for decays in which final state is comprised neutral particles. In the case of $D^0$ decays, indirect CP violation is measured to be  consistent with zero~\cite{Aesp}. So for decays such as $D^0\to\pi^0\pi^0$ and $D^0\to K_S^0\pi^0$, one is effectively measuring the direct CP asymmetry. Within the standard model (SM), mixing and CP violation in the charm sector are expected to be very small \cite{iibigi, gisidori, jbrod}, and thus they constitute a sensitive probe for potential new physics (NP) contributions.

\section{Belle Detector}
The dataset used in these analyses is recorded with the Belle detector at or near various $\Upsilon(nS)$ (n=1,2,3,4,5) resonances. The Belle detector~\cite{Detector} is a general-purpose, large-solid-angle magnetic spectrometer placed around an interaction point (IP) of the KEKB $e^+e^-$ asymmetric-energy collider. It is composed of a tracking system  comprising a silicon vertex detector (SVD) and a 50-layer central drift chamber (CDC), a particle identification (PID) system consisting of an array of aerogel Cherenkov counters (ACC) and time-of-flight scintillation counters (TOF), and an electromagnetic calorimeter made out of CsI(Tl) crystals. All these components are located inside a 1.5 T magnetic field provided by a superconducting magnet. Resistive plate chambers equip the iron flux-return of the magnet, placed at the outermost part of the detector, to identify $K_L$ mesons and muons (KLM).

\section{Experimental Method}    
We select $D^0$ candidates coming from the decay $D^{*+}\to D^0\pi^+_s$, where $\pi^+_s$ denotes the low-momentum (``slow") pion. The charge of this slow pion actually reveals the flavor content of neutral $D$ meson (whether it is a $D^0$ or a $\bar{D}^0$) at its production vertex~\cite{CConj}. A stringent selection criterion is applied on the momentum of the $D^{*+}$ candidate in the $e^+e^-$ center-of-mass frame, $p^*(D^*)$, to suppress $D^{*+}$'s coming from $B$ decays as well as to reduces the combinatorial background. 

The $D^0$ proper decay time is calculated as $t=m_{D^{0}}\vec{L}.\vec{p}/|p|^2$, where $m_{D^0}$ is the invariant mass, $\vec{L}$ is the vector joining the production and decay vertex, and $\vec{p}$ is the momentum of the $D^0$ candidate, respectively. The proper decay time distribution is parametrized as a convolution of an exponential and a resolution function, $\mathcal{R}$, which models the lifetime resolution of the detector. The resolution function, described as a sum of several Gaussian functions, is determined from simulated data samples with some parameters directly coming from the real data. 
  
Charged track candidates are selected by applying requirements on the impact parameters $|dz|$ and $dr$, defined as the distance of closest approach with respect to the IP along the axis of the detector and in the plane perpendicular to it. Later they are identified as kaons or pions based on a likelihood ratio, $\mathcal{P}_{K}\equiv \mathcal{L}_{K}/(\mathcal{L}_{K}+\mathcal{L}_{\pi})$, where $\mathcal{L}_{K}$ ($\mathcal{L}_{\pi}$) is the likelihood of a track being a kaon (pion)~\cite{KID}. Similarly $\mathcal{P}_{\pi}\equiv \mathcal{L}_{\pi}/(\mathcal{L}_{K}+\mathcal{L}_{\pi})$ is the likelihood of a charged track being a pion.
  
%%D0-\barD0  mixing in D0->K+pi-

\section{$D^0$-$\bar{D}^0$ Mixing in $D^0\to K^+\pi^-$}

This study is performed using a 976 fb$^{-1}$ data sample and reports the first observation of $D^0$-$\bar{D}^0$ mixing from an $e^+e^-$ collider~\cite{BRKo}. We define $D^{0}\to K^+\pi^-$ as the wrong-sign (WS) decay while $D^0\to K^-\pi^+$ as the right-sign (RS) process, and measure the ratio of their time-dependent decay to study the mixing phenomenon. The RS amplitude is a sum of the cabibbo-favored (CF) amplitude and a negligibly small contribution from the process of $D^0$-$\bar{D}^0$ mixing followed a doubly cabibbo suppressed (DCS) decay. On the other hand, the WS process is contributed by two comparable amplitudes of a DCS decay and the $D^0$-$\bar{D}^0$ mixing followed by a CF decay. Assuming CP conservation and mixing parameters to be small ($|x|,|y|\ll~1$), the ratio of the WS to RS decay rates, $R(\tilde{t}/\tau)$, is given by
\begin{equation}
\label{Rratio}
R(\tilde{t}/\tau)=\frac{\Gamma_{WS}(\tilde{t}/\tau)}{\Gamma_{WS}(\tilde{t}/\tau)}\approx R_{D}+\sqrt{R_{D}}y'\left(\frac{\tilde{t}}{\tau}\right)+\frac{x'^2+y'^2}{4}\left(\frac{\tilde{t}}{\tau}\right)^2,
\end{equation}
where, $\tilde{t}$ is the true proper decay time, $R_{D}$ is the ratio of the DCS to CF decay rate, and $\tau$ is the $D^0$ lifetime. $x'$ and $y'$ are related to the mixing parameters $x$ and $y$ as, $x'=x\cos\delta+y\sin\delta$, and $y'=y\cos\delta-x\sin\delta$, where $\delta$ is the strong phase difference between the DCS and CF decay.

The phenomenon of mixing can be measured by using Eq. (\ref{Rratio}) only when the measured proper decay time is approximately equal to the true proper decay time. This condition is hardly satisfied in $e^+e^-$ $B$-factories due to a relatively poorer lifetime resolution achieved. Instead, we measure the time-dependent ratio of the WS to RS decay rates as:
\begin{equation}
R(t/\tau)=\frac{\int_{-\infty}^{+\infty}\Gamma_{\text{WS}}(\tilde{t}/\tau)\mathcal{R}(t/\tau-\tilde{t}/\tau)d(\tilde{t}/\tau)}{\int_{-\infty}^{+\infty}\Gamma_{\text{RS}}(\tilde{t}/\tau)\mathcal{R}(t/\tau-\tilde{t}/\tau)d(\tilde{t}/\tau)},
\end{equation}
\label{}
where $t$ is the measured proper decay time and $\mathcal{R}$ is the resolution function of the true proper decay time.

We require the charged tracks to originate from the IP with $dr<$~2 cm, $|dz|<$~4 cm, and to have a transverse momentum greater than 100~MeV/c. The observable $p^*(D^*)$ is constrained to be greater than 2.5, 2.6, and 3.0 GeV/c for the data collected below, at, and above the $\Upsilon(4S)$ resonance, respectively. We select $K$ with $\mathcal{P}_{K}>$~0.4, $\pi$ with $\mathcal{P}_{K}<$~0.7,  $\pi_s$ candidate with $\mathcal{P}_{K}<$~0.9, and the $K\pi$ invariant mass, $M(K\pi)$, within $\pm$20~MeV/c$^2$ of the nominal $D^0$ mass~\cite{Ks_inv_m}. The misidentification of RS decays as WS processes and vice versa are eliminated by applying stringent selections, $\mathcal{P}_{K}>$~0.99 for $K$ and $\mathcal{P}_{K}<$~0.01 for $\pi$, and $M(K\pi)$ lies within $\pm$25~MeV/c$^2$ of the known $D^0$ mass~\cite{Ks_inv_m}. With these requirements we obtained 2980710$\pm$1885 RS and 11478$\pm$177 WS signal candidates.

\begin{figure}[htbp]
\begin{center}
\includegraphics[width=0.47\textwidth]{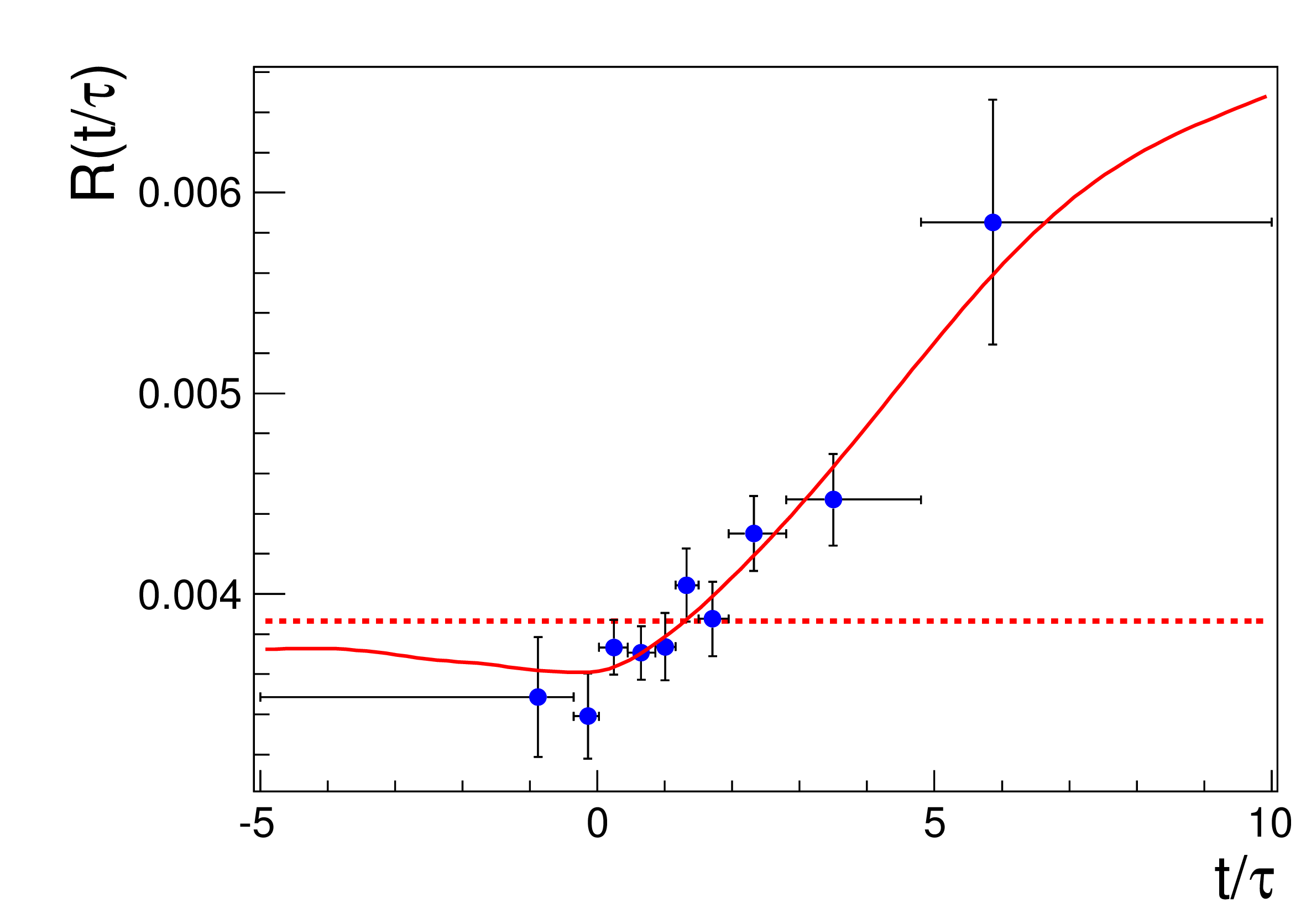}
\caption{(color online).~Time-dependent ratio of the WS to RS process. Points with error bars are the data. The solid (dashed) red curve shows the fit with mixing (no mixing) hypothesis}
\label{fig:Rt}
\end{center}
\end{figure}

We divide the whole data sample into ten bins of proper decay time, and calculate the WS to RS decay-rate ratio as shown in Fig.~\ref{fig:Rt}. This time-dependent ratio is fitted with two hypotheses, with and without mixing, and the results are listed in Table~\ref{tab:summary}.

\begin{table}
\caption{Results of the time-dependent fit to $R(t/\tau)$, where DOF stands for the number of degrees of freedom. The quoted errors are statistical and systematics combined.}

\label{tab:summary}
\begin{tabular}{lcc}
    \hline\hline
    {Test}&{}&{~~~~~~~~~~~Fit}\\
    {hypothesis}&{}&{~~~~~~~~~~~results}\\
    {($\chi^2$/DOF)}&{~~~~~~~~~Parameters}&{~~~~~~~~~~~(10$^{-3}$)}\\
   \hline
    Mixing&~~~~~~~~~R$_D$&~~~~~~~~~~~3.53$\pm$0.13\\
    (4.2/7)&~~~~~~~~~$y'$&~~~~~~~~~~~4.6$\pm$3.4\\
    &~~~~~~~~~$x'^2$&~~~~~~~~~~~0.09$\pm$0.22\\
    \hline
    No mixing&~~~~~~~~~R$_D$&~~~~~~~~~~3.864$\dot{\pm}$0.059\\
    (33.5/9)& &\\
    \hline\hline
    \medskip
\end{tabular}
\end{table}

In conclusion, we report the first observation of $D^0$-$\bar{D}^0$ mixing in $e^+e^-$ collisions with mixing parameters $x'^{2}=(0.09\pm0.22)\times10^{-3}$ and  $y'=(4.6\pm3.4)\times10^{-3}$. The no mixing hypothesis is excluded at the 5.1 standard deviation ($\sigma$) level.

\section{Time-dependent Dalitz Analysis of $D^0\to K^0_S\pi^+\pi^-$}

In this study, we perform a simultaneous measurement of $D^0$-$\bar{D}^0$ mixing and CP violation through a time-dependent Dalitz plot (DP) analysis of the self-conjugate decay $D^0\to K^0_S\pi^+\pi^-$. This is a unique mode with which one can directly determine $x$ and $y$ by studying the distribution of events across the DP as a function of the proper decay time. The results are based on a 921~fb$^{-1}$ of data sample collected with the Belle detector~\cite{TPeng}. 

Assuming the CP symmetry, decay rate for $D^0$($\bar{D}^0$) decaying to the final state $f$ ($\equiv K^0_S\pi^+\pi^-$), with an amplitude $\mathcal{A}_{f}$ ($\mathcal{A}_{\bar{f}}$), is given as:

\begin{align}
|\mathcal{M}(f,t)|^2 & = \frac{e^{-\Gamma t}}{2} \left\{\left(|\mathcal{A}_{f}|^2+\left|\frac{q}{p}\right|^2 |\mathcal{A}_{\bar{f}}|^2\right)\cosh(\Gamma y t) \right.\nonumber\\
 &\quad \left. {} + \left(|\mathcal{A}_{f}|^2-\left|\frac{q}{p}\right|^2 |\mathcal{A}_{\bar{f}}|^2\right)\cos(\Gamma x t) \right. \nonumber\\
 &\quad \left. {} +2\mathfrak{R}\left(\frac{q}{p}\mathcal{A}_{\bar{f}}\mathcal{A}_{{f^{*}}}\right)\sinh(\Gamma y t) \right. \nonumber\\
 &\quad \left. {} - 2\mathfrak{J}\left(\frac{q}{p}\mathcal{A}_{\bar{f}}\mathcal{A}_{f^{*}}\right)\sin(\Gamma x t)\right\}, {\text {and}}
\end{align}

\begin{align}
|\mathcal{\bar{M}}(f,t)|^2 & = \frac{e^{-\Gamma t}}{2} \left\{\left(|\mathcal{A}_{\bar{f}}|^2+\left|\frac{p}{q}\right|^2 |\mathcal{A}_{f}|^2\right)\cosh(\Gamma y t) \right.\nonumber\\
 &\quad \left. {} + \left(|\mathcal{A}_{\bar{f}}|^2-\left|\frac{p}{q}\right|^2 |\mathcal{A}_{f}|^2\right)\cos(\Gamma x t) \right. \nonumber\\
 &\quad \left. {} +2\mathfrak{R}\left(\frac{p}{q}\mathcal{A}_{f}\mathcal{A}_{{\bar{f^{*}}}}\right)\sinh(\Gamma y t) \right. \nonumber\\
 &\quad \left. {} - 2\mathfrak{J}\left(\frac{p}{q}\mathcal{A}_{f}\mathcal{A}_{\bar{f}^{*}}\right)\sin(\Gamma x t)\right\}.
\end{align}

The decay rates are exponential in the $D^0$ lifetime, $\tau_{D^0}=1/\Gamma$, and are modulated by the mixing parameters $x$ and $y$. The mixing parameters can be extracted from an unbinned maximum likelihood fit of these decay rates to the DP distribution as a function the $D^0$ decay time.

\begin{figure}[htbp]
\begin{center}
\begin{tabular}{cc}
\includegraphics[width=0.225\textwidth]{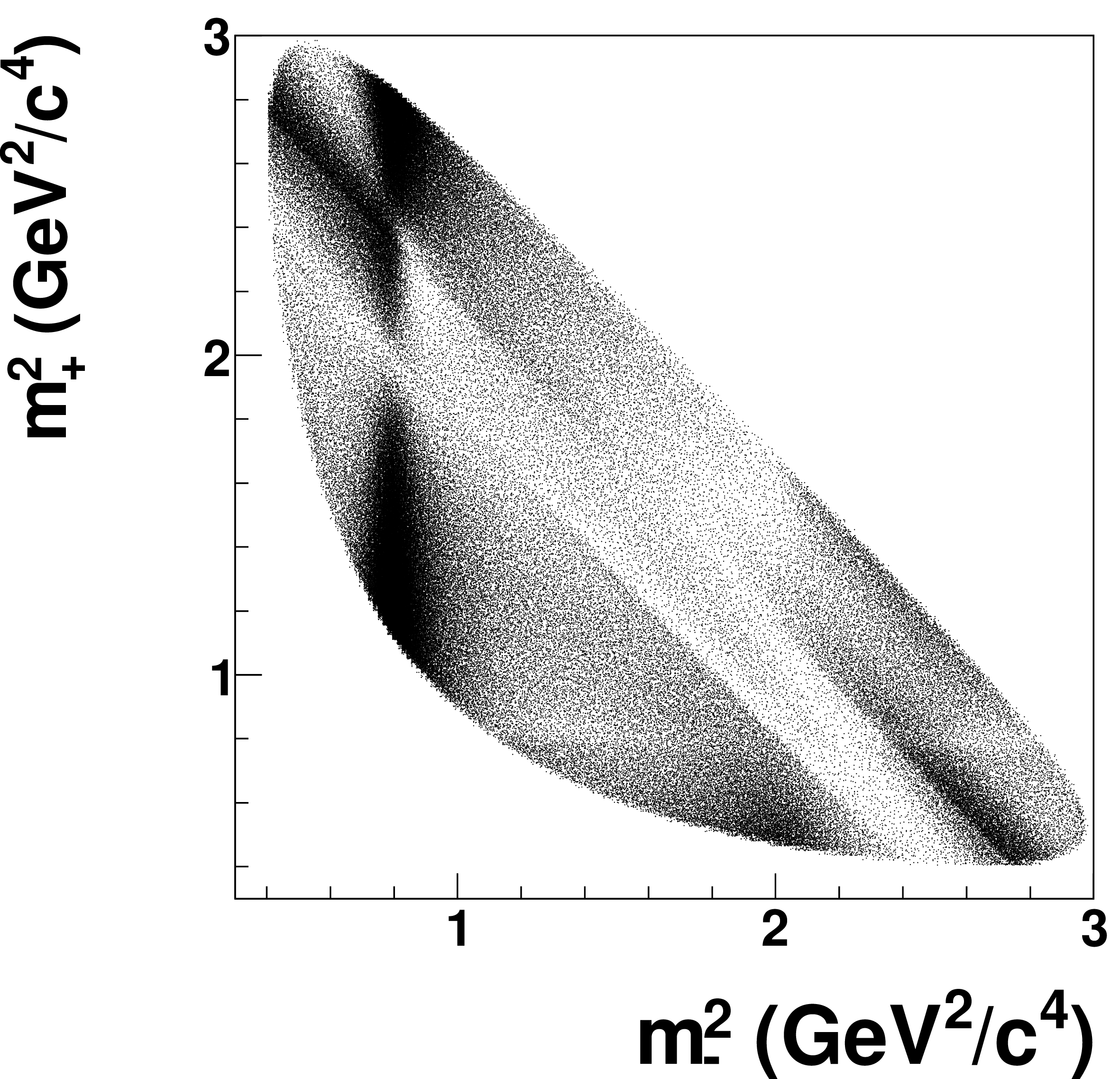}
\includegraphics[width=0.225\textwidth]{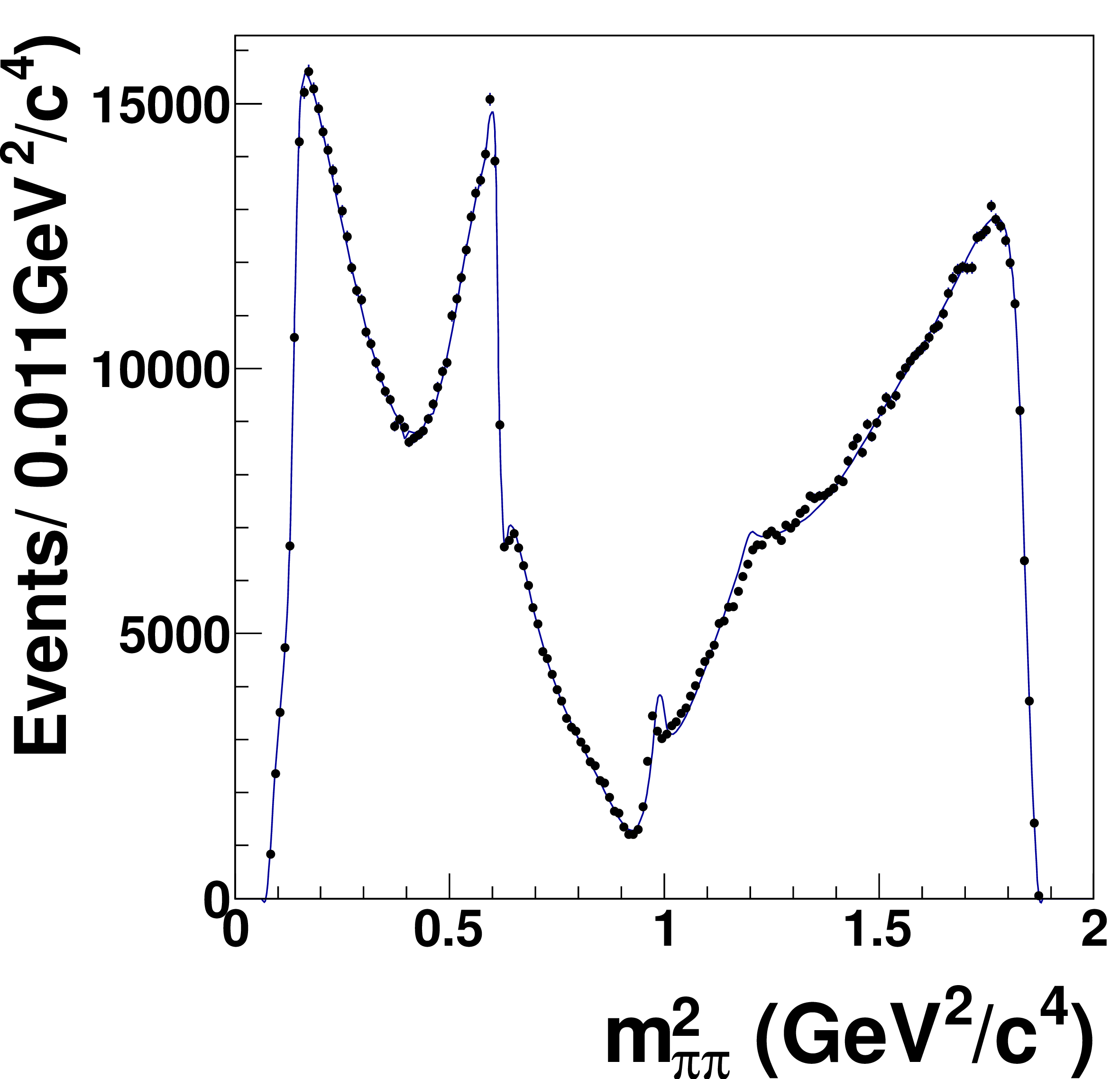}

\end{tabular}
\begin{tabular}{cc}
\includegraphics[width=0.225\textwidth]{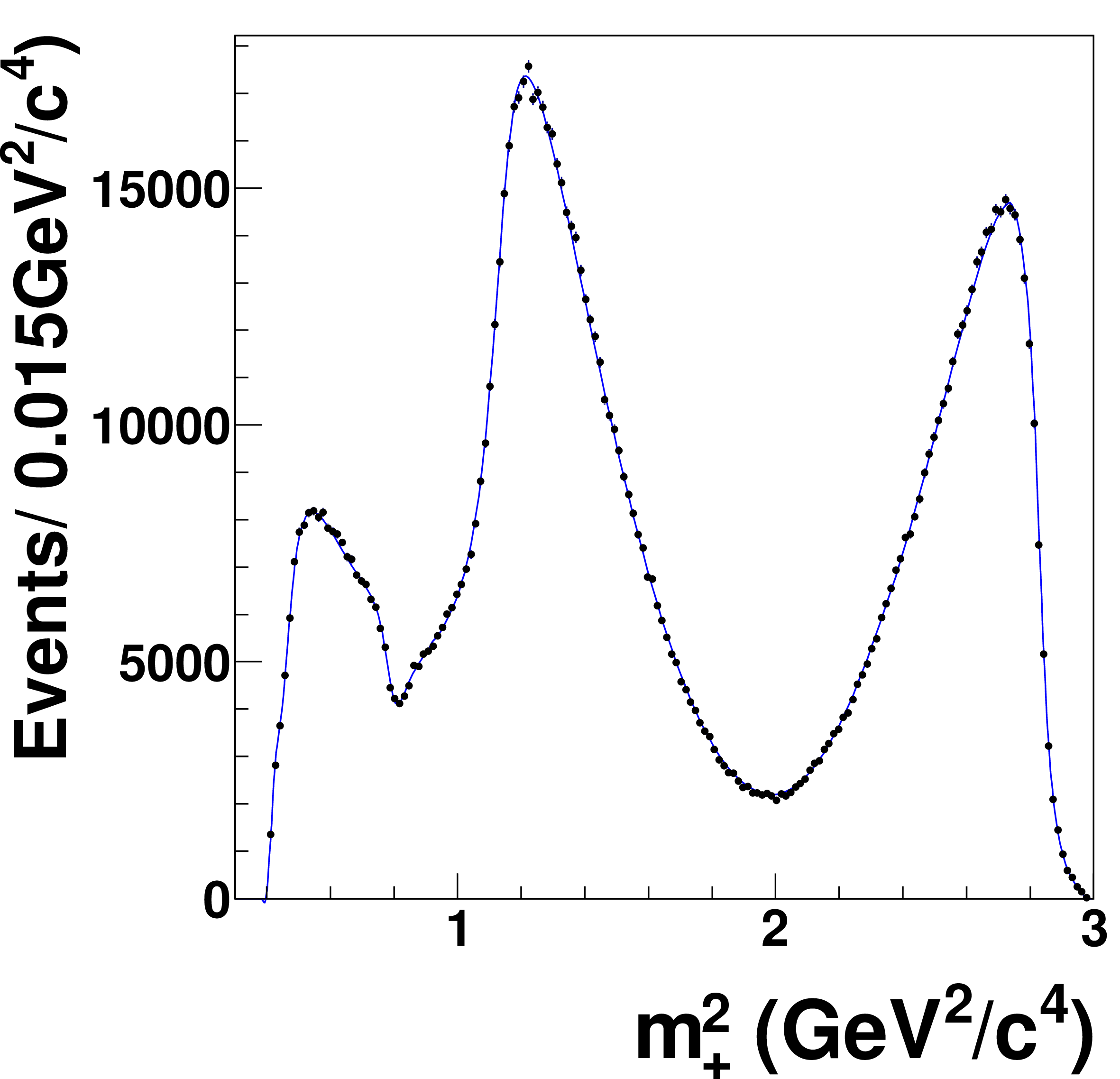}
\includegraphics[width=0.225\textwidth]{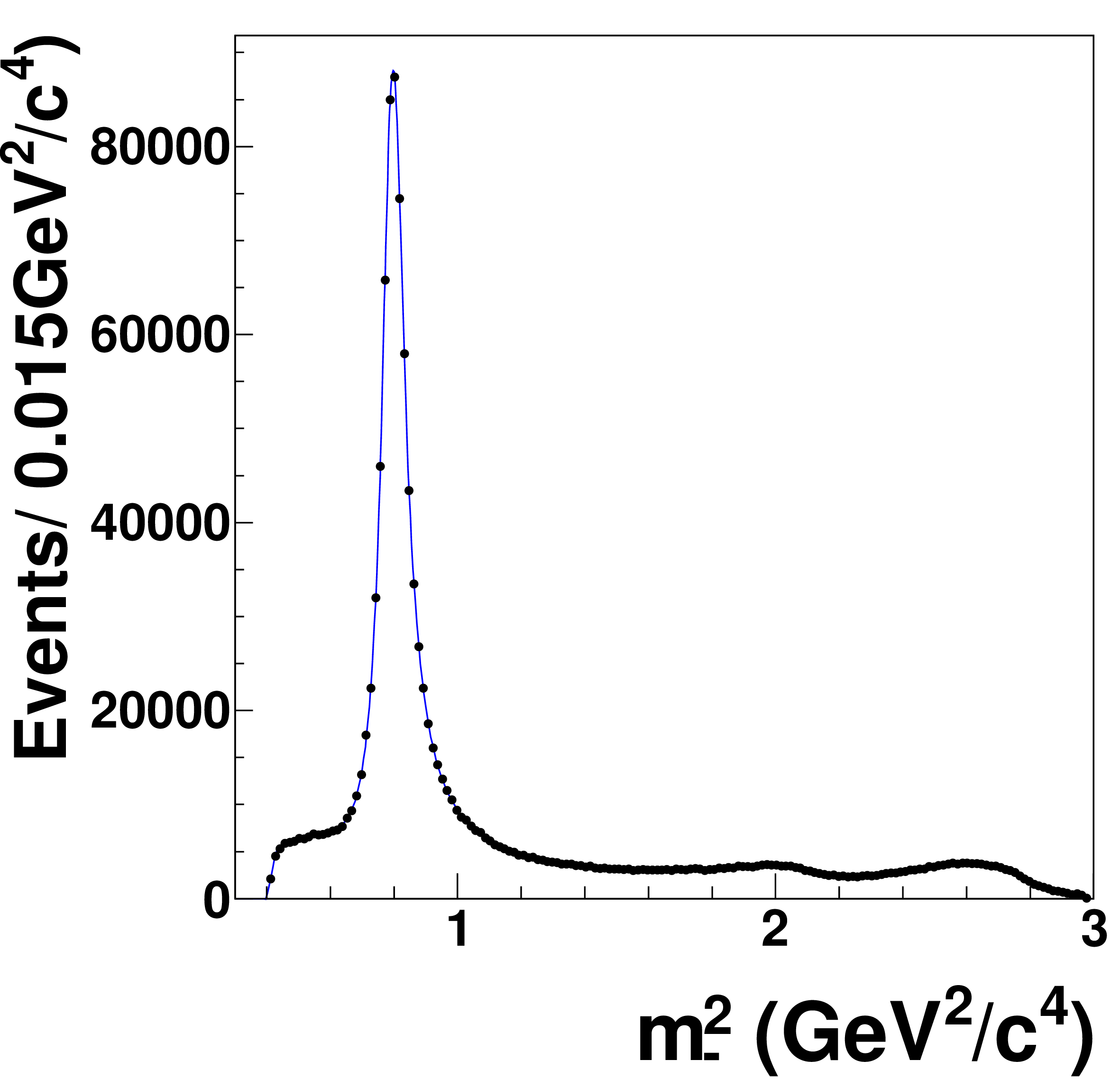}

\end{tabular}
\caption{(color online). Dalitz distribution and projections of Dalitz variables $m_+^2$, $m_-^2$, and $m_{\pi\pi}^2$. Points are data with error and solid lines are fit result.}
\label{fig:DalitzP}
\end{center}
\end{figure}

The $D^0$ candidate is reconstructed by combining a $K_S^0$ with two oppositely charged pions that have at least two SVD hits and an invariant mass lying within about $\pm$10 MeV/c$^2$ of the nominal $K^0_S$ mass. We require $p^*(D^*)$ to be greater than 2.5 and 3.1 GeV/c for $\Upsilon(4S)$ and $\Upsilon(5S)$ samples, respectively.
Events with $D^0$ invariant mass ($M_{K_S^0\pi^+\pi^-}$) ranging from 1.81  to 1.92 GeV/c$^2$ and the kinetic energy released during the $D^*$ decay (Q) falling within a range (0, 20) MeV are selected. We obtain the signal yield by a two dimensional fit comprising $M_{K_S^0\pi^+\pi^-}$ and $Q$, and find 1231731$\pm$1633 candidates with a purity of 95.5\%. The selected events are shown in Fig.~\ref{fig:DalitzP} as a function of DP variables. 

%%Dalitz plot

As the three body decay $D^0\to K^0_S\pi^+\pi^-$ can proceed via many quasi-two-body intermediate states involving resonances, the Dalitz amplitudes $\mathcal{A}_{f}$ and $\bar{\mathcal{A}}_{f}$ are expressed as a sum of quasi-two-body amplitudes. Twelve intermediate states with resonances of P- and D-wave dynamics are modeled by a relativistic Breit-Wigner parametrization. For the $\pi\pi$ S-wave we use the K-matrix formalism \cite{KM} while for the $K\pi$ S-wave we adopt the LASS model \cite{LASS}.

A time integrated fit to DP is performed separately for $D^0$ and $\bar{D}^0$ and allows to assume that $\mathcal{\bar{A}}_f$ is equal to $\mathcal{A}_f$. In subsequent fit we assume CP is conserved, keep the mixing parameters $x$, $y$ and $D^0$ lifetime, $\tau$, are free. We obtain mixing parameters $x=(0.56\pm0.19)\%$, $y=(0.30\pm0.15)\%$  and $\tau=(410.3\pm0.6)$~fs. We perform a CPV search by performing fit while CPV parameters $|q/p|$ and arg$(q/p)$ are free in addition to the mixing parameters and lifetime. The values of mixing parameters obtained are identical to the CP-conserved case, and the CPV parameters are $|q/p|=0.90^{+0.16}_{-0.15}$ and arg$(q/p)=(-6\pm11)^{\circ}$, consistent with no CP violation.

%%D0->Pi0Pi0
\section{Time-integrated CP asymmetry in $D^0\to\pi^0\pi^0$}

Even though mixing has been observed in charm sector, evidence for the CP violation is not reported yet. SM expectation of CP violation is of the order of 0.1\%. In 2012, LHCb~\cite{LHCb} and CDF~\cite{CDF} have measured a large CP asymmetry difference between $D^0\to K^+ K^-$ and $D^0\to \pi^+ \pi^-$, $\Delta A_{CP}$, resulting in a value 4.6$\sigma$~\cite{HFAG} away from zero . Isospin consideration can relate the $\Delta A_{CP}$ to CP  asymmetries in decay modes like $D^0\to\pi^0\pi^0$~\cite{IsoTheory}.
In this study we measure time-integrated CP asymmetry in $D^0\to\pi^0\pi^0$ and update the existing measurement for $D^0\to K_S^0\pi^0$, using a data sample of  integrated luminosity 966~fb$^{-1}$~\cite{Nis}.

A reconstruction asymmetry is measured 
\begin{equation}
A_{\text{rec}}=\frac{N_{\text{rec}}^{D^{*+}\to D^0\pi^+_s}-N_{\text{rec}}^{D^{*-}\to \bar{D}^0\pi^-_s}}{N_{\text{rec}}^{D^{*+}\to D^0\pi^+_s}+N_{\text{rec}}^{D^{*-}\to \bar{D}^0\pi^-_s}},
\end{equation}
where $N_{\text{rec}}$ is the number of reconstructed signal events. The $A_{\rm rec}$ comprises three contributions: the underlying CP asymmetry $A_{CP}$, the forward-backward asymmetry, and detection efficiency asymmetry $A_{\epsilon}^{\pi_{s}}$ arising due to difference in the detection efficiency of $\pi^+$ and $\pi^-$ inside the detector. The $A_{FB}$ is a function of production angle of $D^*$ candidate in $e^+e^-$ center of momentum frame ($\theta^*$) while $A_{\epsilon}^{\pi_{s}}$ depends on the transverse momentum $p_{T}^{\pi_s}$ and polar angle $\theta^{\pi_s}$ of $\pi_{s}$. The difference in reconstruction asymmetry of $D^0\to K^-\pi^+$ (``untagged") and $D^*+\to D^0\pi^+_s\to K^-\pi^+\pi^+_s$ (``tagged") gives $A_{\epsilon}^{\pi_{s}}$ by assuming same $A_{FB}$ for $D^{*+}$ and $D^0$ mesons \cite{Aesp}. After correcting for $A_{\epsilon}^{\pi_{s}}$, the reconstruction asymmetry becomes
\begin{equation}
A_{\text{rec}}^{\text{cor}}=A_{CP}+A_{FB}(\cos\theta^*),
\end{equation}
The $A_{CP}$ is independent of any kinematic variable while $A_{FB}$ is an odd function of cosine of $\theta^*$. The $A_{CP}$ and $A_{FB}$ are therefore extracted by dividing the data samples into ten bins of $\cos\theta^*$ and later by adding or subtracting $A_{\text{rec}}^{\text{cor}}$ values belonging to the bins with same value of $|\cos\theta^*|$.
\begin{equation}
A_{CP}=[A_{\text{rec}}^{\text{cor}}(\cos\theta^*)+A_{\text{rec}}^{\text{cor}}(-\cos\theta^*)]/2,
\label{eq:Acp}
\end{equation}
and
\begin{equation}
A_{FB}=[A_{\text{rec}}^{\text{cor}}(\cos\theta^*)-A_{\text{rec}}^{\text{cor}}(-\cos\theta^*)]/2.
\label{eq:Afb}
\end{equation}

The $K_S^0$ candidates are reconstructed from two oppositely charged pions with an invariant mass lying within 9 MeV/c$^2$ of nominal $K_S^0$ mass \cite{Ks_inv_m}, and $\pi^0$'s are reconstructed from two $\gamma$'s where the invariant mass lies within 25 MeV/c$^2$ of $\pi^0$ mass \cite{Ks_inv_m}. The $K_S^0$ and $\pi^0$ candidates are combined to form the $D^0$ candidate with invariant mass lying within (1.750, 1.950) GeV/c$^2$ and two $\pi^0$'s  combined to form $D^0$ candidate with an invariant mass within (1.758. 1.930) GeV/c$^2$. We require that the $\pi_s$ candidate originates from IP with $dr<$~1 cm and $|dz|<$~3 cm and with $\mathcal{P}_{\pi}>$~0.4. With these selection criteria we obtain the signal yield of 34460$\pm$273 for $D^0\to\pi^0\pi^0$ and 466814$\pm$773 for $D^0\to K_S^0\pi^0$.

\begin{figure}[htbp]
\begin{center}
\begin{tabular}{cc}
\includegraphics[width=0.22\textwidth]{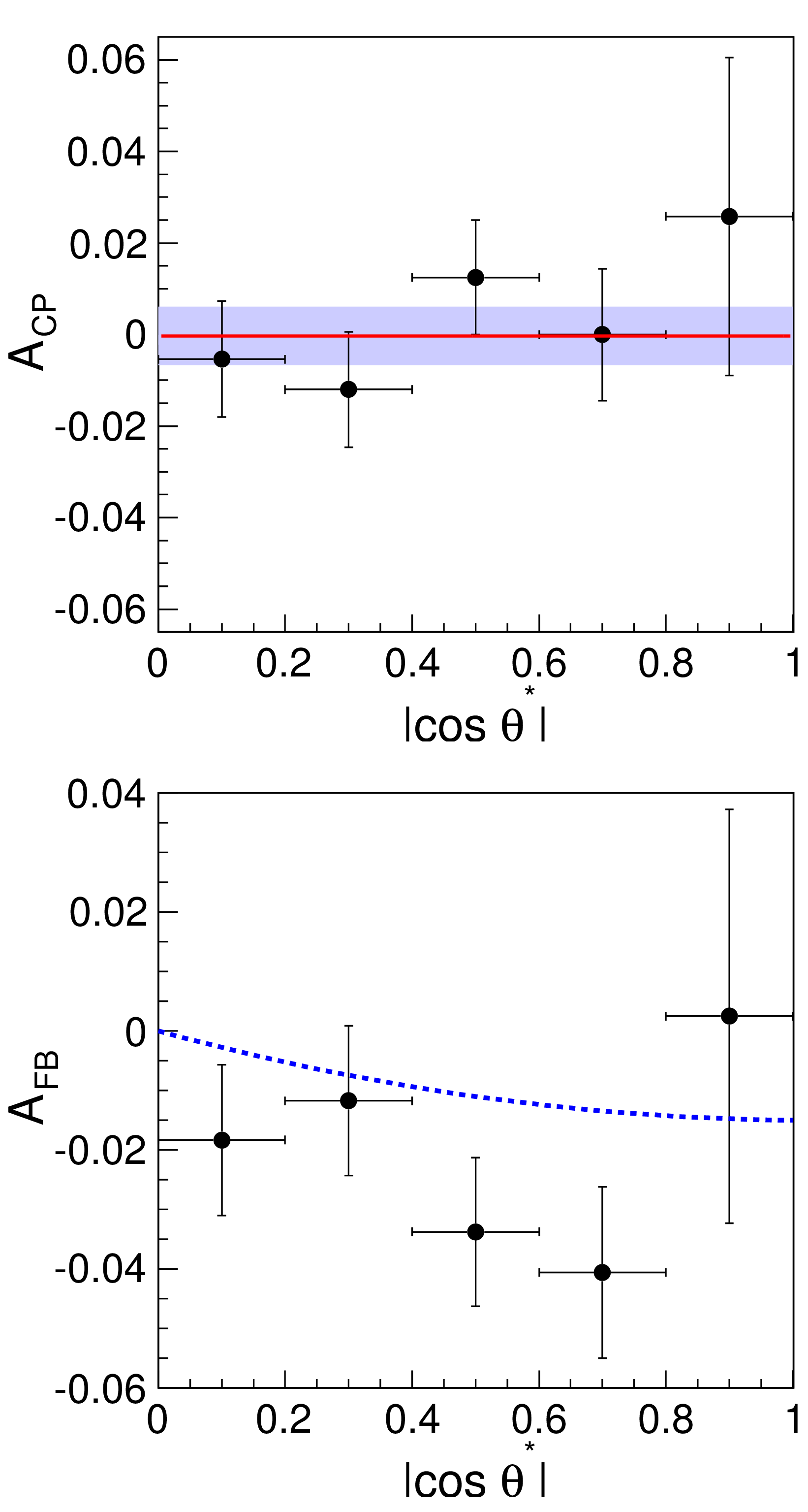}
\includegraphics[width=0.22\textwidth]{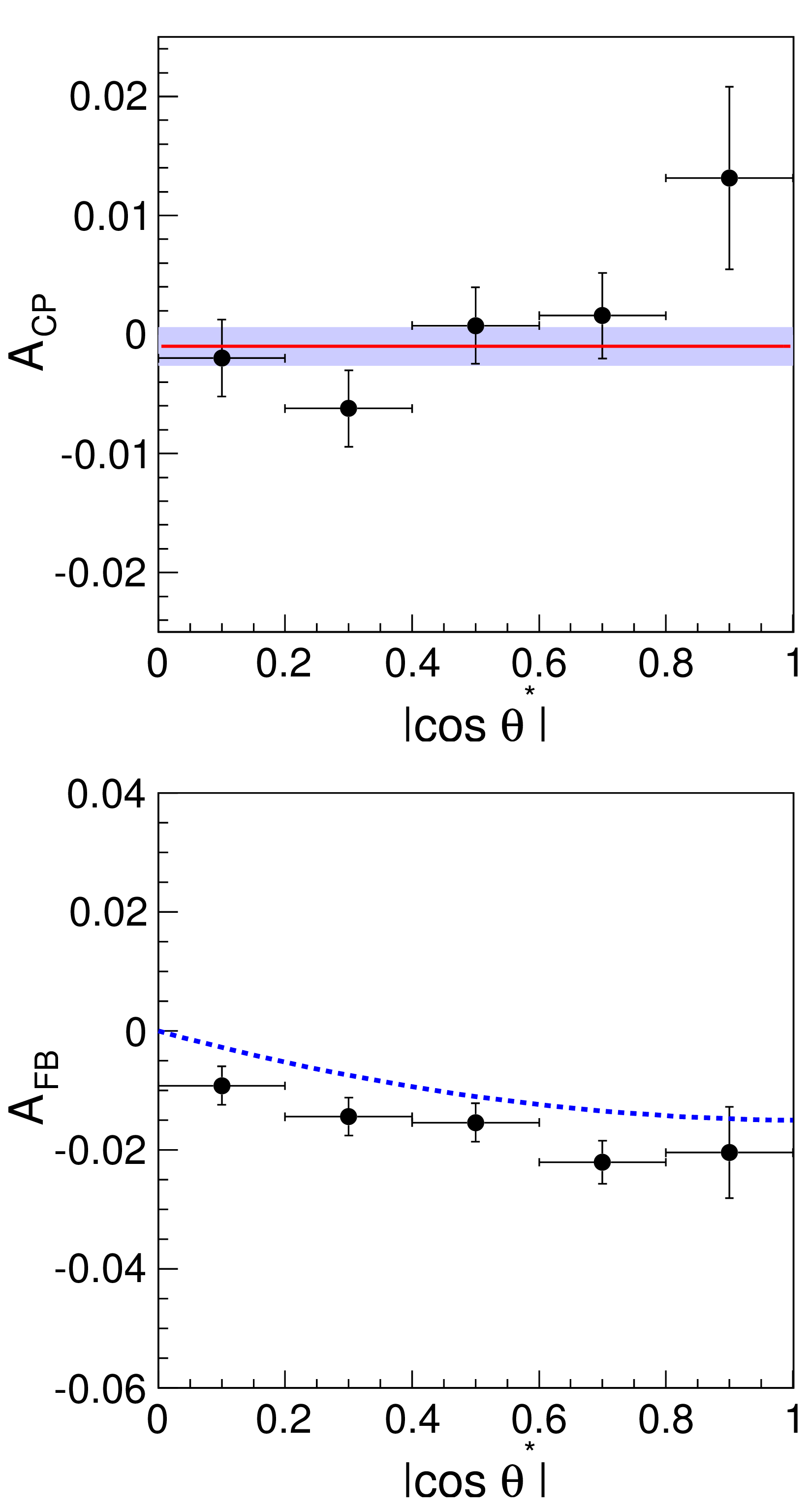}

\end{tabular}
\caption{(color online). CP asymmetry $A_{CP}$ (top) and forward-backward asymmetry as a function of $|\cos\theta^*|$. Plots on left (right) side are for the $\pi^0\pi^0$ ($K_S^0\pi^0$) final states. The solid red lines are the central value, the blue band shows 1$\sigma$ interval, and the dashed blue curve for the $A_{FB}$ plots shows leading order prediction for $A_{FB}$.}
\label{fig:acp_kspi_pipi}
\end{center}
\end{figure}

The SM predicts a non-zero CP asymmetry in the decays containing neutral kaons in the final state, which arises due to $K^0$-$\bar{K}^0$ mixing, and the expected magnitude is $A_{CP}^{\bar{K}^0}=(-0.339\pm 0.007)\%$~\cite{KKmix}.
The selected data sample is first divided in ten bins of $\cos\theta^*$, and the content of each bin is further divided into 7$\times$8 bins of $[p_T^{\pi_s}, \cos\theta^{\pi_s}]$. A simultaneous fit~\cite{Nis} is performed to get the asymmetries ($A_{\text{rec}}$). Each of $A_{\text{rec}}$ is corrected for $A_{\epsilon}^{\pi_s}$, obtained in Ref.~\cite{Effmap}, to get $A_{\text{rec}}^{\text{cor}}$. A weighted average of $A_{\text{rec}}^{\text{cor}}$ over $[p_T^{\pi_s}, \cos\theta^{\pi_s}]$ bins having same value of $\cos\theta^*$ is performed, and $A_{CP}$ and $A_{FB}$ are extracted from Eqs.~(\ref{eq:Acp}) and (\ref{eq:Afb}) respectively. Fig.~\ref{fig:acp_kspi_pipi} shows $A_{CP}$ and $A_{FB}$ as a function of $|\cos\theta^*|$ for both $\pi^0\pi^0$ and $K_S^0\pi^0$ final state.

%%Acp plot

By fitting and applying corrections we obtained $A_{CP}(\pi^0\pi^0)=(-0.03\pm0.64)\%$ and   $A_{CP}(K_S^0\pi^0)=(-0.10\pm0.16)\%$. The systematic uncertainties are estimated as in Ref.~\cite{Nis}, adding the effect $A_{CP}^{\bar{K}^0}$ for $A_{CP}(K_S^0\pi^0)$, and adding the systematic uncertainty, we obtaine the final result $A_{CP}(\pi^0\pi^0)=(-0.03\pm0.64\pm0.10)\%$ and $A_{CP}(K_S^0\pi^0)=(-0.21\pm0.16\pm0.07)\%$. These results are consistent with no CP violation and significantly improve over the previously published results.

\section{Conclusion}
In conclusion, we report the first observation of $D^0$-$\bar{D}^0$ mixing in $e^+e^-$ collision by measuring a time-dependent ratio of WS to RS decay rates and the no mixing hypothesis is excluded at 5.1$\sigma$ level. A time-dependent DP analysis is performed for $D^0\to K_S^0\pi^+\pi^-$, the measurement of mixing is updated  with a significance of 2.5$\sigma$ away from no-mixing point, and no evidence for CP violation is observed. We report a significantly improved measurement of time integrated CP violating asymmetry in $D^0\to\pi^0\pi^0$ and the result is consistent with no CP violation. Also we report an updated measurement of CP asymmetry in $D^0\to K_S\pi^0$ decays.
\nocite{*}
\bibliographystyle{elsarticle-num}
\bibliography{martin}

\begin{thebibliography}{00}
\bibitem{CConj} The charge conjugation is applied throughout unless stated explicitly.
\end{thebibliography}

\end{document}